\documentclass[%
notitlepage,
twocolumn, reprint,
superscriptaddress,
amsmath,amssymb,
aps,
pra,
floatfix,
]{revtex4-1}

\usepackage{amsmath}
\usepackage{graphicx}
\usepackage[T1]{fontenc}
\usepackage{hyperref}   
\usepackage[export]{adjustbox}
\usepackage{enumitem}
\usepackage{xcolor}
\definecolor{magenta}{rgb}{1.0,0.0,1.0}
\definecolor{green}{rgb}{0.0,0.5,0.0}
\definecolor{gray}{gray}{0.9}

\usepackage[normalem]{ulem}

\begin{document}
\title{Combining machine learning with physics:\\A framework for tracking and sorting multiple dark solitons}
\author{Shangjie Guo}
\affiliation{Joint Quantum Institute, National Institute of Standards and Technology, and University of Maryland, Gaithersburg, Maryland 20899, USA}
\author{Sophia M. Koh}
\affiliation{Department of Physics and Astronomy, Amherst College, Amherst, Massachusetts 01002, USA}
\affiliation{National Institute of Standards and Technology, Gaithersburg, Maryland 20899, USA}
\author{Amilson R. Fritsch}
\affiliation{Joint Quantum Institute, National Institute of Standards and Technology, and University of Maryland, Gaithersburg, Maryland 20899, USA}
\author{I.~B.~Spielman}
\affiliation{Joint Quantum Institute, National Institute of Standards and Technology, and University of Maryland, Gaithersburg, Maryland 20899, USA}
\author{Justyna P. Zwolak}
\email{jpzwolak@nist.gov}
\affiliation{National Institute of Standards and Technology, Gaithersburg, Maryland 20899, USA}

\date{\today}
\begin{abstract}
In ultracold-atom experiments, data often comes in the form of images which suffer information loss inherent in the techniques used to prepare and measure the system.
This is particularly problematic when the processes of interest are complicated, such as interactions among excitations in Bose-Einstein condensates (BECs).
In this paper, we describe a framework combining machine learning (ML) models with physics-based traditional analyses to identify and track multiple solitonic excitations in images of BECs. 
We use an ML-based object detector to locate the solitonic excitations and develop a physics-informed classifier to sort solitonic excitations into physically motivated subcategories.
Lastly, we introduce a quality metric quantifying the likelihood that a specific feature is a longitudinal soliton.
Our trained implementation of this framework, \textsc{soldet}, is publicly available as an open-source \textsc{python} package. 
\textsc{soldet} is broadly applicable to feature identification in cold-atom images when trained on a suitable user-provided dataset.
\end{abstract}
\maketitle

\section{Introduction}

Machine learning (ML) techniques promise improved data analysis and enhanced performance for today's quantum devices and technologies.
Ultracold atomic gases are a nearly ideal system to deploy ML-driven analysis, where the automated exploration and interpretation of a very large dataset, in the form of images, can lead to scientific enhancements and experimental optimization~\cite{Mei20-OQF} as well as new discoveries. 
Here we focus on the general problem of feature identification, a commonly recurring task in the analysis of such data, from locating vortices~\cite{Abo-Shaeer2001,Schweikhard2004,Lin2009b} or tracking solitons~\cite{Becker2008,Aycock2017}, identifying spin textures or magnetic domain walls~\cite{Sadler2006a,Vengalattore2010,De2014} to locating topological singular points~\cite{Tarnowski2017}.
While data from these examples have been individually analyzed using task-specific algorithms (or even manual inspection), they are all feature identification problems that can be solved using a single ML-enhanced analysis framework.  
This paper introduces such a framework, and demonstrates its utility on the specific problem of identifying solitonic excitations in atomic Bose-Einstein condensates (BECs), as well as quantifying the quality of each identified feature.

Traditional statistical analysis using physics-based models, such as least-square fitting and hypotheses testing, have been go-to techniques for data analysis since the 1800's~\cite{Legendre1805} and remain widely applied in quantum cold-atom image analysis~\cite{ketterle1999making,fritsch_creating_2020, purdy2017quantum}.
The outcome of physics-model-based algorithms and fits are intuitive, physically meaningful, and can help identify patterns present in the data; even fits based on more heuristic functions can have coefficients that are derived in obvious ways from the data. 
By contrast, ML methods work as ``black boxes,'' making their operation difficult to interpret.
Conventional statistical methods use fixed algorithms in conjunction with preconceived models for data reduction.
Overfitting occurs when the number of fit parameters is comparable or larger than the number of independent data points.
In this context, the process of training an ML tool essentially codesigns the fitting algorithm and the data model, as encoded by a large number of internal parameters.
Training ML models is itself a fitting process that can be susceptible to overfitting, for example when the training dataset has too little variability or the ML model has too many internal parameters.
ML involves a class of data-driven techniques that do not rely on preexisting models, but also add additional opportunities for overfitting that can make them less reliable on new data than conventional techniques.

\begin{figure*}[t]
    \includegraphics[width=0.83\linewidth]{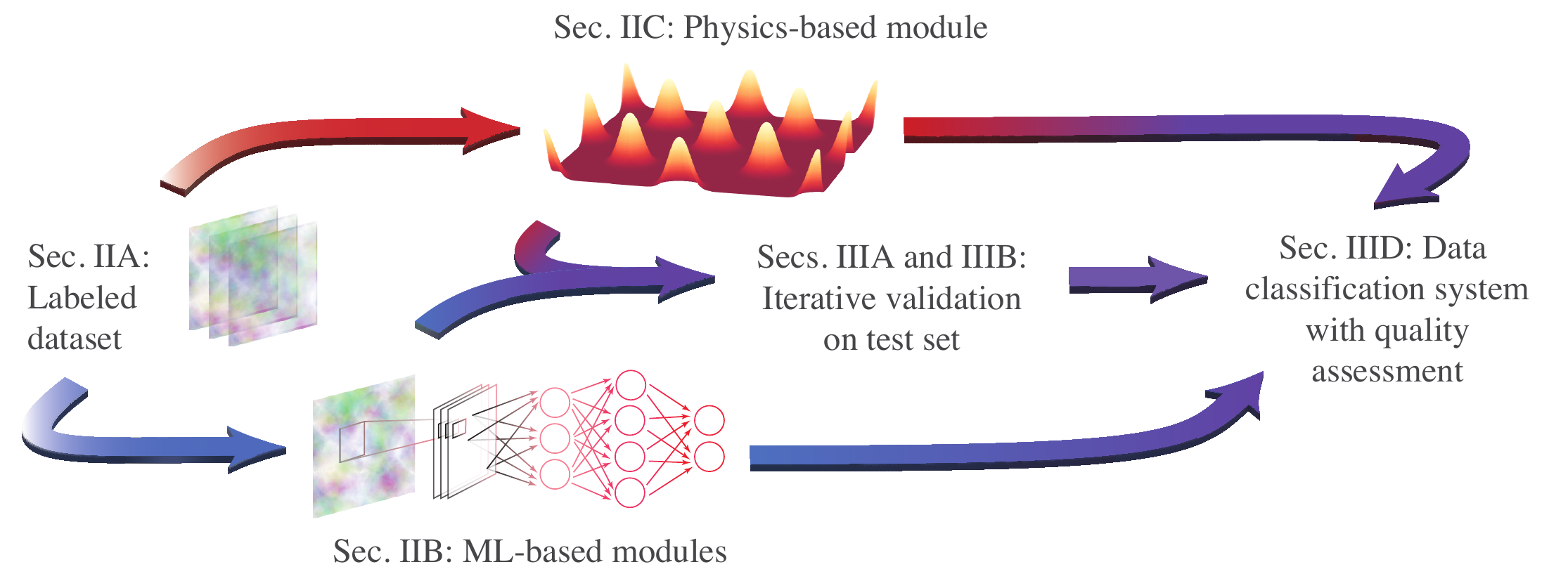}
    \caption{Overview of the framework. The colored arrows link the preparation (Secs.~\ref{ssec:data}, \ref{ssec:ML}, and \ref{ssec:physics_based}), validation (Secs.~\ref{ssec:od_test} and \ref{ssec:ec_test}), and application (Sec.~\ref{ssec:test}) phases of the framework.
    The red path represents the preparation and implementation of the physics-based-approximation module of the framework.
    The blue path represents the ML modules.}
    \label{fig:framework}
\end{figure*}

Here, we describe the hybrid two-module feature identification framework shown in Fig.~\ref{fig:framework}, that combines the flexibility of ML techniques with the intuition and robustness of conventional fitting methods.
Furthermore the separate outputs of these two very different modules allow us to assess data quality by cross-validation. 
Hybrid approaches have been employed in other settings, for example for landslide prediction~\cite{huang2020comparisons}, medical image processing~\cite{ghosh2012new}, and cyber attack detection~\cite{sakhnini2019smart}.

The framework begins with a labeled dataset that is used to train the ML module and initialize the physics-based module.
Before trusting either module, we independently validate each module on a subset of the labeled data that was not used for training.
Model redesign may be needed until satisfactory performance of each module is reached.
We then combine both modules into an integrated system able to analyze new data.

We demonstrate the performance of our framework using data from atomic BECs, quintessential quantum systems.
Quantum research with BECs, and cold-atom quantum gases more broadly, is multifaceted with examples ranging from realizing collective many-body physics~\cite{Greiner2002} to creating today's most accurate atomic clocks~\cite{Bothwell2019}.
In the vast majority of these experiments, data is acquired in the form of noisy images that typically have undergone evolution, such as a time of flight, before measurement.
This often obfuscates the computation of the quantities of interest.
Cold quantum gases therefore make an ideal testbed for our methodology that combines physically motivated, but heuristic, fitting functions with established computer vision techniques.

\begin{figure*}[!hb]
    \vspace{5pt}
    \centering
    \includegraphics[width=0.88\linewidth]{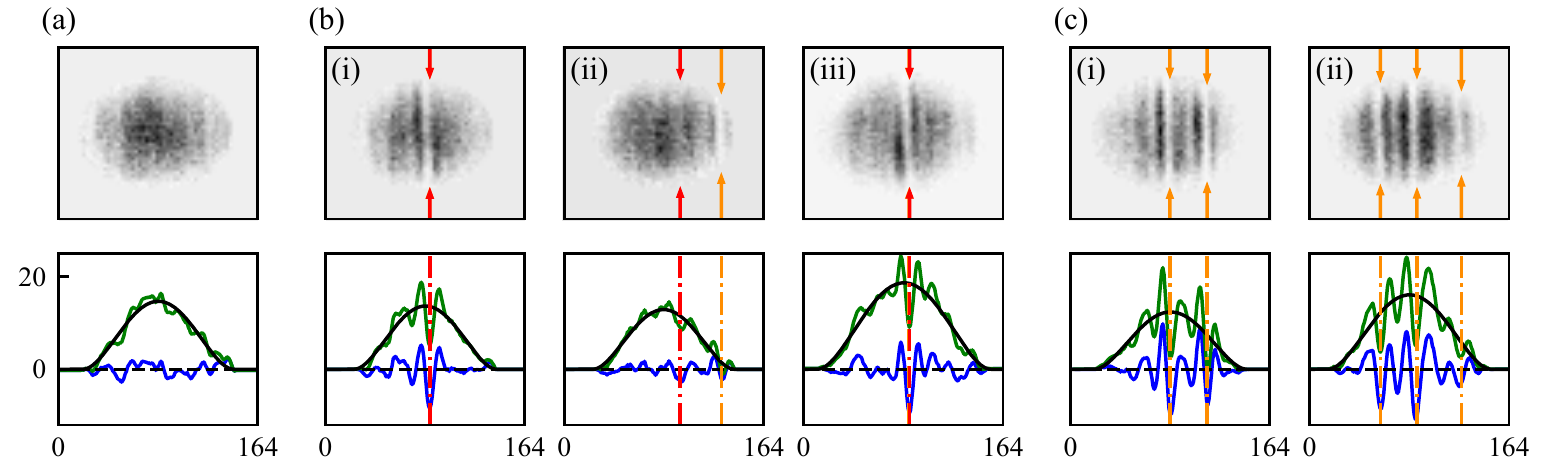}
    \caption{Representative data.
    The top panels plot preprocessed images from our dataset and the bottom panels plot profiles: profile of full image (green), TF fits (black), density fluctuations (blue).
    The red lines mark the location of the deepest depletion in the density fluctuations, while the orange lines mark the soliton locations found from our OD.
    (a) An element of the no-excitation class.
    (b) Three elements of the single-excitation class: (i) a single longitudinal soliton, (ii) an off-center longitudinal soliton, and (iii) a solitonic vortex.
    (c) Two representative elements of the other excitations class.
    }
    \label{fig:sample_data}
\end{figure*}

We focus on the specific problem of locating dark solitons (spatially compact excitations that manifest as reductions in the atomic density) as they move in BECs~\cite{burger_dark_1999,Denschlag2000,fritsch_creating_2020}.
This allows us to leverage our established soliton dataset~\cite{guo_machine-learning_2021,solitons-data} to train and validate our framework; representative elements of the dataset are shown in Fig.~\ref{fig:sample_data}.
These data consist of elliptical atom clouds (top row) where solitons appear as vertically aligned density depletions (bottom row).
Not all vertically aligned density depletions are created equal: deep depletions mark the location of slowly moving kink solitons; shallow depletions are associated with rapidly moving kink solitons or ``longitudinal'' solitonic vortices (where the vortex core is aligned in the image plane); asymmetric depletions can result from ``transverse'' solitonic vortices~\cite{Mateo2015} (where the vortex core is aligned perpendicularly to the image plane); and chains of stripes can result from highly excited phonon modes.
Our framework is a tool that can automatically locate all the solitonic excitations in each image and distinguish between longitudinal solitons and transverse solitonic vortices.
Here we introduce the term ``longitudinal soliton'' to include both kink solitons and longitudinal solitonic vortices.

Our ML module leverages and extends established computer vision techniques.
Computer vision is a broad field with applications ranging from image classification to semantic segmentation and object detection~\cite{voulodimos2018deep}.
Object detection refers to the capability of software systems to locate and identify objects in an image. 
Convolutional 

\onecolumngrid

\twocolumngrid
\noindent neutral networks (CNNs) underlie solutions to all of these tasks, and unsurprisingly were employed in our previous work classifying soliton image data into three categories: no solitonic excitation, one solitonic excitation, and other excitations~\cite{guo_machine-learning_2021}.
Our ML module goes beyond simple classification and uses a CNN-based object detector (OD) to provide the location of all candidate excitations in a given image.

By contrast our physics-based module employs a least-squares fit of an inverted and skewed Mexican-hat function to one-dimensional (1D) background-subtracted projections of soliton candidates (shown in bottom row in Fig.~\ref{fig:sample_data}).
We initialized this module using our previously labeled single soliton data and employ a Yeo-Johnson transformation~\cite{yeo_new_2000} to produce a multivariate normal distribution yielding the likelihood that an unknown feature is a soliton.

This approach yielded three immediate benefits.
First, a careful analysis of the coefficients from the physics based-module identified previously overlooked correlations that allow us to distinguish between some solitonic excitations (longitudinal solitons and transverse solitonic vortex~\cite{burger_dark_1999,Denschlag2000,donadello_observation_2014,Mateo2015}).
Second, combining the results of the ML and fitting modules allowed us to automatically create a larger, more reliable dataset that includes fine-grained information such as the soliton position and type of excitation.
This dataset is described in Ref.~\cite{Fritsch21-DSD} and published in the NIST data repository~\cite{solitons-data}. 
Third, our hybrid framework was prepared solely from a training dataset whose images contain either zero or one solitonic excitation; however, it is performant on complex data containing multiple excitations.

The remainder of this paper is structured as follows: 
Section~\ref{sec:modules} introduces both modules and describes their training and initialization.
Section~\ref{sec:result} describes the validation of both modules and their performance on new data that include multiple solitonic excitations.
In Sec.~\ref{ssec:soldet}, we describe an open-source \textsc{python} reference implementation of our framework: \textsc{soldet}~\cite{SolDet}.
Lastly, in Sec.~\ref{sec:discussion} we conclude and discuss the potential applications of the framework as well as the possible future directions. 

\section{Data and modules}\label{sec:modules}
In addition to the recent success of ML methods~\cite{ metz_deep_2020,guo_machine-learning_2021,leykam2021dark}, solitonic excitations have also been located and characterized using traditional fitting techniques.
For example, Ref.~\cite{fritsch_creating_2020} began with the background-removed atom density profiles (blue curves in Fig.~\ref{fig:sample_data}) described in Sec.~\ref{ssec:data}, then identified the deepest depletion (orange dashed line), and fit to a Gaussian function (a physically motivated, but heuristic choice) centered near the deepest depletion. 
This yielded physical information including soliton width, depth, and position. 
Unfortunately, this simple approach is failure prone, as for example in Fig.~\ref{fig:sample_data}(b)(ii), where the deepest depletion is far from the actual soliton. 
Moreover, it detects only single solitonic features, making human intervention necessary when many excitations are present.
Rather than finding the deepest minimum, our framework first uses an OD (described in Sec.~\ref{ssec:ML}) to provide an initial estimate of all solitonic excitation positions, and then uses a skewed Mexican-hat fit function (Sec.~\ref{ssec:physics_based}) that accurately describes their density profiles.
The resulting fit coefficients serve two purposes: qualitative likelihood assessment and fine-grained categorization.

\subsection{Data}\label{ssec:data}
Our framework is trained and initialized using a revised dataset consisting of about $5.5 \times 10^3$ manually labeled experimental images of BECs with and without solitonic excitations~\cite{Fritsch21-DSD,solitons-data}.
The experimental setup and preprocessing techniques are described in~\cite{fritsch_creating_2020}.

Figure~\ref{fig:sample_data} shows six selected sample images from the labeled dataset.
The dataset includes labels for five classes:
``no solitonic excitation,'' images that do not contain any excitations; ``single solitonic excitation,'' images containing one solitonic excitation; ``other excitations,'' images not in the preceding classes (including those with multiple solitonic excitations, high degrees of noise, and those annotators could not agree on); ``mislabeled'', data determined to be potentially mislabeled during the curation process; and ``unlabeled,'' images that have not been manually annotated.
Additionally, for the single excitation class the dataset includes the horizontal position of excitations within BEC.

Figure~\ref{fig:sample_data}(a) displays an image from the no excitation class, which lacks the pronounced stripes present in the remaining examples. 
In (b), we show three elements of the single excitation class, each containing a single dark vertical fringe: (b)(i) a longitudinal soliton; (b)(ii) an off-center single longitudinal soliton; and (b)(iii) a solitonic vortex. 
In (c), we show two elements of the other excitations class containing more than one vertical fringe.

Horizontal 1D profiles (bottom row of Fig.~\ref{fig:sample_data}) also have features associated with vertically aligned solitonic excitations and are amenable to least-squares fitting.
We obtain these profiles by first summing the pixel values vertically to compress two-dimensional (2D) images to 1D; this sum can be over all (green curves) or part (see Sec.~\ref{sssec:pie_class}) of the vertical extent of the image.
We then fit a 1D Thomas-Fermi (TF) model
\begin{eqnarray}
  n^{\rm TF}(i) &= \, n_0\, {\max}\left\{\left[1-\left(\frac{i-i_0}{R_0}\right)^2\right],0\right\}^{2} + \delta_n
  \label{eqn:cloudbg}
\end{eqnarray}
to each summed 1D profile, where $i$ is the horizontal pixel index, and $n_0$, $i_0$, $R_0$, and $\delta_n$ are fitting parameters representing peak density, center position, TF radius, and an overall offset, respectively.
This fit (black curves) serves as an overall background that we subtract from the 1D profiles, leaving behind the 1D density fluctuations (blue curves). 
The orange dashed lines represent the location of deepest depletion in the 1D fluctuations.

\subsection{ML modules}\label{ssec:ML}
Our previous dark soliton classifier~\cite{guo_machine-learning_2021} consisted of a CNN model that returned one of the three predefined classes: no solitonic excitation, single solitonic excitation, or other excitations.
However, this detector did not locate the excitations.
To compare with experimental data, we located the soliton by identifying the deepest depletion and fitting to a Gaussian, as described above.
This algorithm has two limitations: 
(1) The soliton may not be the deepest depletion [as in Fig.~\ref{fig:sample_data}(b)(ii)];
and (2) multiple solitons cannot be located [as in Fig.~\ref{fig:sample_data}(c)].
Here we retain the CNN classifier to globally organize the data, but inspired by a highly successful recent result using an OD to locate vortices in numerically simulated 2D BECs~\cite{metz_deep_2020}, we employ an OD to locate solitonic excitations in experimental images of highly elongated BECs.

The OD complements the CNN classifier in two ways: (1) it identifies soliton positions rather than classifying; and (2) even though it is trained with single-soliton data, it can locate multiple excitations in the same image.
We employ a neural network based OD with six convolution layers and four max-pooling layers but no fully connected layers (see the Appendix~\ref{app:ml} for more details). 
The OD has an order of magnitude fewer trainable parameters than our previous CNN ($7\times10^4$ versus $\sim 10^6$ parameters), accelerating the training process and making it lightweight to deploy.
Because the OD simply requires a dataset with many representative instances of the object to be detected, it requires far less training data than the CNN classifier (which by design required substantial data from all considered classes).

In our data, the solitonic excitations are roughly four pixels in width.
Since our images are $164$ pixels wide, we designed our OD to aggregate the image into $41$ spatial cells, each with two outputs in the range $[0,1]$; the OD therefore returns a $41 \times 2$ array $\widetilde{\bf Y}$.
For our dataset this aggregation guarantees that each output cell can describe the state of at most one soliton. 
$\widetilde{Y}_{\ell,1}$ is a probability estimate that cell $\ell$ contains a soliton, and $\widetilde{Y}_{\ell,2}$ is the fractional position of the soliton center within that cell, where $0$ or $1$ correspond to the left or right edge of the cell, respectively.
The OD considers any cell with $\widetilde{Y}_{\ell,1}>0.5$ as containing an excitation, and then obtains its position from $\widetilde{Y}_{\ell,2}$.

When comparing to the training dataset with labels denoted by ${\bf Y}$, we use the cost function~\cite{metz_deep_2020} 
\begin{align}\label{eqn:cost}
   F = \sum_{\ell=1}^{41} 
   \begin{cases}
     -w_1  \log(\widetilde{Y}_{\ell,1})+ w_2(Y_{\ell,2} - \widetilde{Y}_{\ell,2})^2,& \text{if } Y_{\ell,1}=1\\
     - \log(1-\widetilde{Y}_{\ell,1}),              & \text{if } Y_{\ell,1}=0
 \end{cases}
\end{align}
for each training image, where the label $Y_{\ell,1}$ identifying the presence of an excitation in a cell is fully confident, i.e., either 0 or 1.
The coefficients $w_1$, $w_2$ are hyperparameters controlling the relative importance of each term.
The logarithmic terms increase the cost function when the OD misidentifies solitons, while the quadratic term contributes when a soliton is mislocated within a cell.
Our training set uses images with at most one soliton, so cells with $Y_{\ell,1}=1$ are much less frequent than those with $Y_{\ell,1}=0$; as a result we expect that $w_1, w_2 \gg 1$ to give similar overall weight to the three terms in Eq.~(\ref{eqn:cost}).
We train the OD by minimizing the cost function summed over all training images, updating the predicted OD values $\widetilde{\bf Y}$ in each iteration.
Because the cell size is comparable to the soliton size, a single soliton can span two cells.
To prevent double counting, we merge detections occurring in adjacent cells and take the position to be their average.
 
We deem the OD's detection successful if our training data contains a labeled soliton close to the detected one (within three pixels in our implementation).
The two failure modes are failing to detect a solitonic excitation and reporting an excitation that is not present.

\subsection{Physics-based modules}\label{ssec:physics_based}
In this section, we introduce our physics-based module that uses constrained least-squares fitting to estimate soliton parameters, and following a Yeo-Johnson transformation~\cite{yeo_new_2000}, produces a quality estimate giving the likelihood of a given feature being solitonic.

We fit the Ricker wavelet~\cite{Ricker43-TWT}, i.e., a Mexican-hat function
\vspace{-10pt}
\begin{eqnarray}
  f(i) =&\ \delta_n - \ n^{\rm TF}(i_c) A \exp\left[-\frac{1}{2}\left(\frac{i-i_c}{\sigma}\right)^2\right]\nonumber \\
  &\ \ \ \times \left[1 - a\left(\frac{i-i_c}{\sigma}\right)^2 + b\left(\frac{i-i_c}{\sigma}\right)\right] ,
  \label{eqn:mexhat}
\end{eqnarray}
to the 1D density fluctuations described Sec.~\ref{ssec:data}, where $n^{\rm TF}(i_c)$ is evaluated with $\delta_n=0$.
The function takes six parameters: normalized logarithmic amplitude $A$, center position $i_c$, width $\sigma$, logarithmic symmetrical shoulder height $a$, asymmetrical shoulder height $b$, and an offset $\delta$.
When $a$ and $b$ are zero this function is a simple Gaussian, making $a$ nonzero adds symmetric shoulders to the distribution, and $b$ introduces an asymmetry.
Our solitonic features are well described by this function; since our excitations manifest as density depletions, the second term in Eq.~(\ref{eqn:mexhat}) is negative.

Our constrained least-squares fit requires initial guesses for all of these parameters.
The guess for the center position $i_c$ also provides the initial guess for $A$ by setting it equal to the 1D density fluctuations evaluated at $i_c$.
We found the initial values $\sigma=4$, $a=0.2$, $b=0$, and $\delta=0$ to lead to convergent fits across the whole dataset.
In order to produce reliable fits we apply the following constraints: $i_c$ must remain within three pixels from the initial guess, $10^{-13} < A < 10^{4}$, and $10^{-13} < a < 10^{4}$ to prevent numerical fitting errors.

\subsubsection{Physics-informed excitation classifier}\label{sssec:pie_class}
Many candidate solitonic excitations are not vertically symmetric as might be expected [see, e.g., Fig.~\ref{fig:sample_data}(b)(iii)].
The location of the largest ``shoulder'' in the top half of the excitation is reversed with respect to the bottom half; in addition, the location of the minimum is slightly displaced going from the top half to the bottom.
Inspired by these differences, we bisect each image into top and bottom halves (labeled by $+$ and $-$, respectively) and separately apply the Mexican-hat fit to fluctuations in these data, giving vectors $\Theta^\pm$.
Using this observation, we develop a physics-informed excitation (PIE) classifier based on the single-soliton dataset and discover that correlations between these vectors allow for a more fine-grained excitation classification.

\begin{figure}[t]
    \centering
    \includegraphics{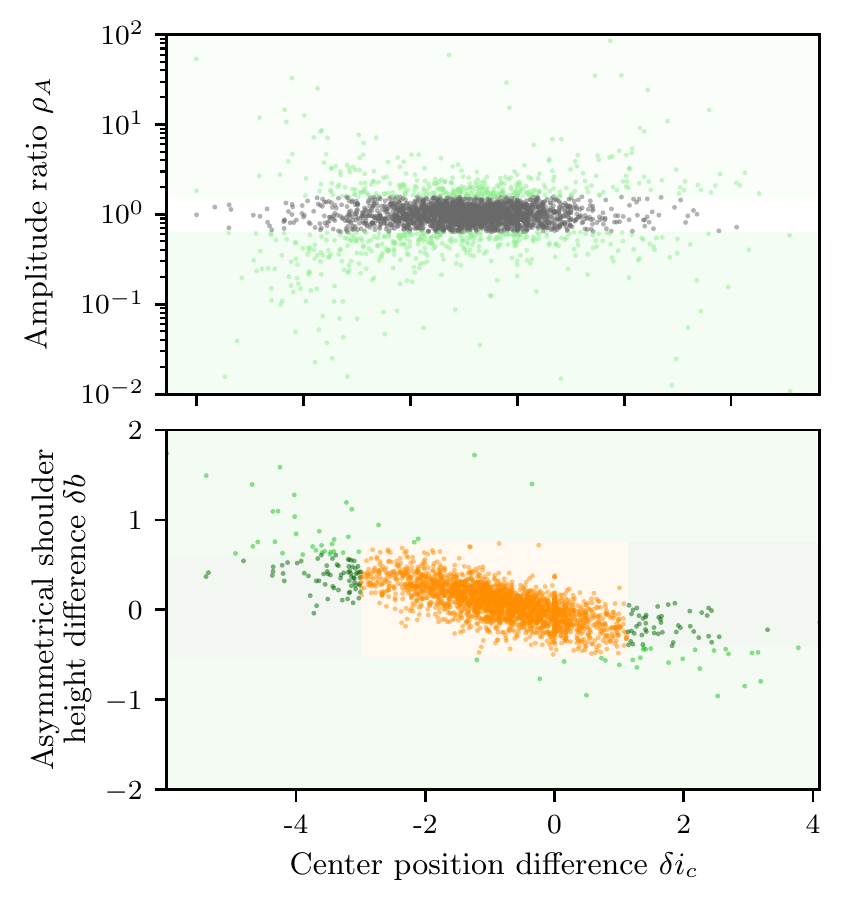}
    \caption{Correlations between parameters implemented in PIE classifier.
    The top panel shows the distribution of center position difference versus the amplitude ratio (on a logarithmic scale).
    The bottom panel shows the correlation between the center position difference and the asymmetrical shoulder height difference for the gray points from the top panel.
    Both panels are colored based on the cut-off points discussed in Sec.~\ref{ssec:ec_test}.}
    \label{fig:parameters}
\end{figure}

Figure~\ref{fig:parameters} shows the distribution of parameters from a single-soliton dataset that were useful for classifying excitations.
No meaningful correlations were found for parameters $\sigma^\pm$ and $a^\pm$, thus these did not assist in classification.
The markers in the top panel show the amplitude ratio $\rho_A = A^+/A^-$ versus the top-bottom position difference $\delta i_c = i_c^+ - i_c^-$, and show that they are not correlated.
By contrast, the bottom panel shows that the asymmetric shoulder height difference $\delta b = b^+/\sigma^+ - b^-/\sigma^-$ is clearly anticorrelated with $\delta i_c$.
Both panels are colored based on the cut-off points discussed in Sec.~\ref{ssec:ec_test} (see also Fig.~\ref{fig:classification_alg}).

This distribution and its correlation guide the classification rules described in Sec.~\ref{ssec:ec_test}, yielding a PIE classifier based on cutoffs defined by human examination of the data. 

\subsubsection{Quality estimation}\label{sssec:qual_est}

Here we describe a quality estimate that a candidate excitation in an image is solitonic. 
We derive the likelihood that a vector of fit outcomes $\Theta = [A, i_c, \sigma, a, b]$ is drawn from a $k=5$ dimensional prior distribution spanning the set of representative solitonic excitations~\footnote{We found that $\delta$ was strongly correlated with the remaining five parameters and did not improve the quality estimate performance}.
Ideally this distribution would be an uncorrelated multivariate normal distribution, but it is not.
As a result, we developed the following procedure to bring the distribution into this desired form.

We first fit a Yeo-Johnson power transformation~\cite{yeo_new_2000} to each separate parameter distribution (having summed the five-dimensional distribution along the remaining parameters) to transform them into independent zero-mean 1D Gaussian distributions with unit variance. 
Note that this treatment cannot transform the parameter distributions into perfect Gaussians; nevertheless, each resulting distribution is balanced, contains a single peak, and has long tails.
The covariance matrix $\Sigma_k$ is uncorrelated after this treatment and the distribution is qualitatively Gaussian in shape.

To calculate the quality estimate for a candidate excitation detected in an image, we
\begin{enumerate}[nosep]
\item fit the subtracted background 1D profile to Mexican-hat function given in Eq.~(\ref{eqn:mexhat}) to obtain $\Theta$;
\item  use the established power transformation on $\Theta$ to obtain $\Theta^\prime$; and
\item  return the quality estimate: $M(\Theta^\prime) = 1 - \chi^2_{k}\left[D^2(\Theta^\prime)\right]$, the likelihood between 0 and 1 that the excitation is solitonic.
\end{enumerate}
The chi-squared cumulative distribution function $\chi^2_k(p)$ relates the Mahalanobis distance~\cite{mahalanobis1936generalized} $D^2(\Theta^\prime) = \Theta^{\prime\dagger} \Sigma^{-1}_k \Theta^\prime$ to the likelihood that an outcome was drawn from the specified distribution~\footnote{This argument assumes no prior knowledge about the distribution of fit outcomes for structures that are not solitonic excitations.}.
$D(\Theta^\prime)$ is unbounded above and decreases to zero as $\Theta^\prime$ approaches $\langle \Theta^\prime\rangle$, the average over the prior distribution.

\section{Results}\label{sec:result}
\subsection{ML modules}\label{ssec:od_test}

\begin{figure}[t]
    \centering
    \includegraphics{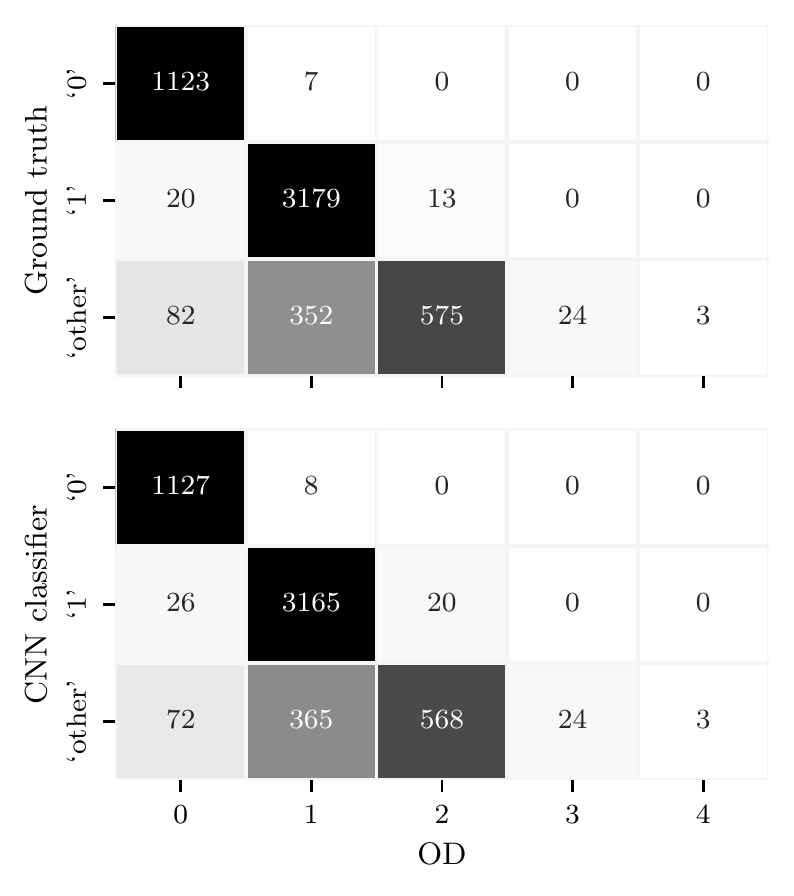}
    \caption{OD performance compared to ground truth (top), and the CNN classifier prediction (bottom). 
    For ground truth and the CNN classifier, the ticks `0,', `1,' and `other' represent no, single, and other excitation classes.
    For OD, ticks represent the total number of positive excitations within an image.
}
    \label{fig:confusion_matrices}
\end{figure}

We train both the CNN classifier and the OD using the refined dataset with added soliton position labels (see Ref.~\cite{Fritsch21-DSD}).
The CNN classifier is trained using the full dataset while the OD training uses only the no solitonic excitation and single solitonic excitation classes.
We assess the performance of both modules using five-fold cross-validation, that is using 80~\% of the data to train a given module and the remaining 20~\% to test it, and repeating the process five times to fully cover the dataset (see the Appendix~\ref{app:ml} for training details).

The results are summarized in the two cumulative confusion matrices plotted in Fig.~\ref{fig:confusion_matrices}.
The top panel compares the outcome of the OD to the initial labels, showing  near perfect delineation between the no excitations and single excitations classes. However, the OD further subdivides the other excitations class, counting anywhere from zero to four candidate solitonic excitations within it.
This results from the existence of excitations in this class that are not solitonic, as well as the possibility of having multiple solitons in the same image.
The analogous comparison to CNN classification labels in the bottom panel is nearly indistinguishable from the one presented in the top panel, evidencing the quality of the CNN predictions. 

Together, these ML tools effectively classify these data and locate excitations; however, they do not provide any fine-grained information on the nature or the quality of the identified excitations.
This is addressed in the following sections.

\subsection{PIE classifier}\label{ssec:ec_test}

The PIE classifier operates by applying a sequence of ``cuts'' driven by different combinations of the top-bottom fit outcomes $\Theta^\pm$.
The exact parameter values described below are arrived at manually by exploring the data accepted and rejected by the cut to minimize the number of false-positive longitudinal soliton identifications.

\begin{figure}[t]
    \centering
    \includegraphics{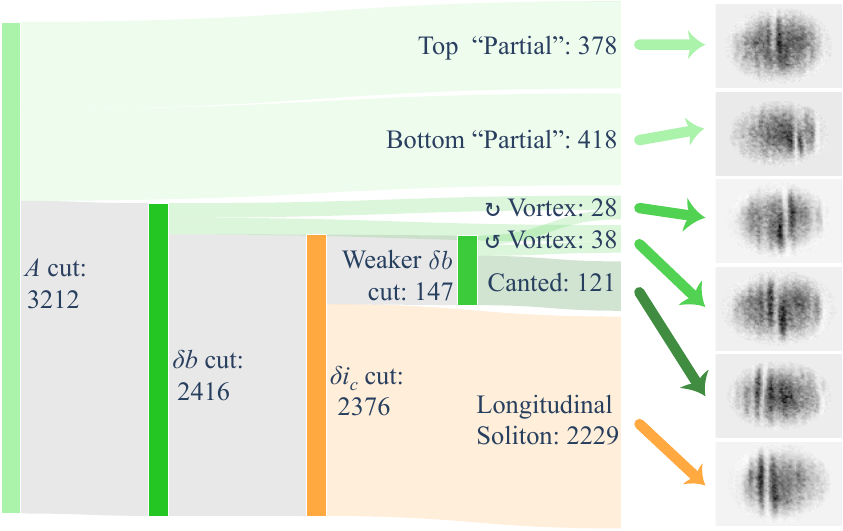}
    \caption{
    The flow of the PIE classifier with example images for classification categories. 
    Flow pathways and nodes are square-root scaled.
    }
\label{fig:classification_alg}
\end{figure}

The following cuts are applied sequentially, and the PIE classifier stops as soon as a classification is assigned.

\begin{figure*}[t]
    \centering
    \includegraphics[width=0.99\linewidth]{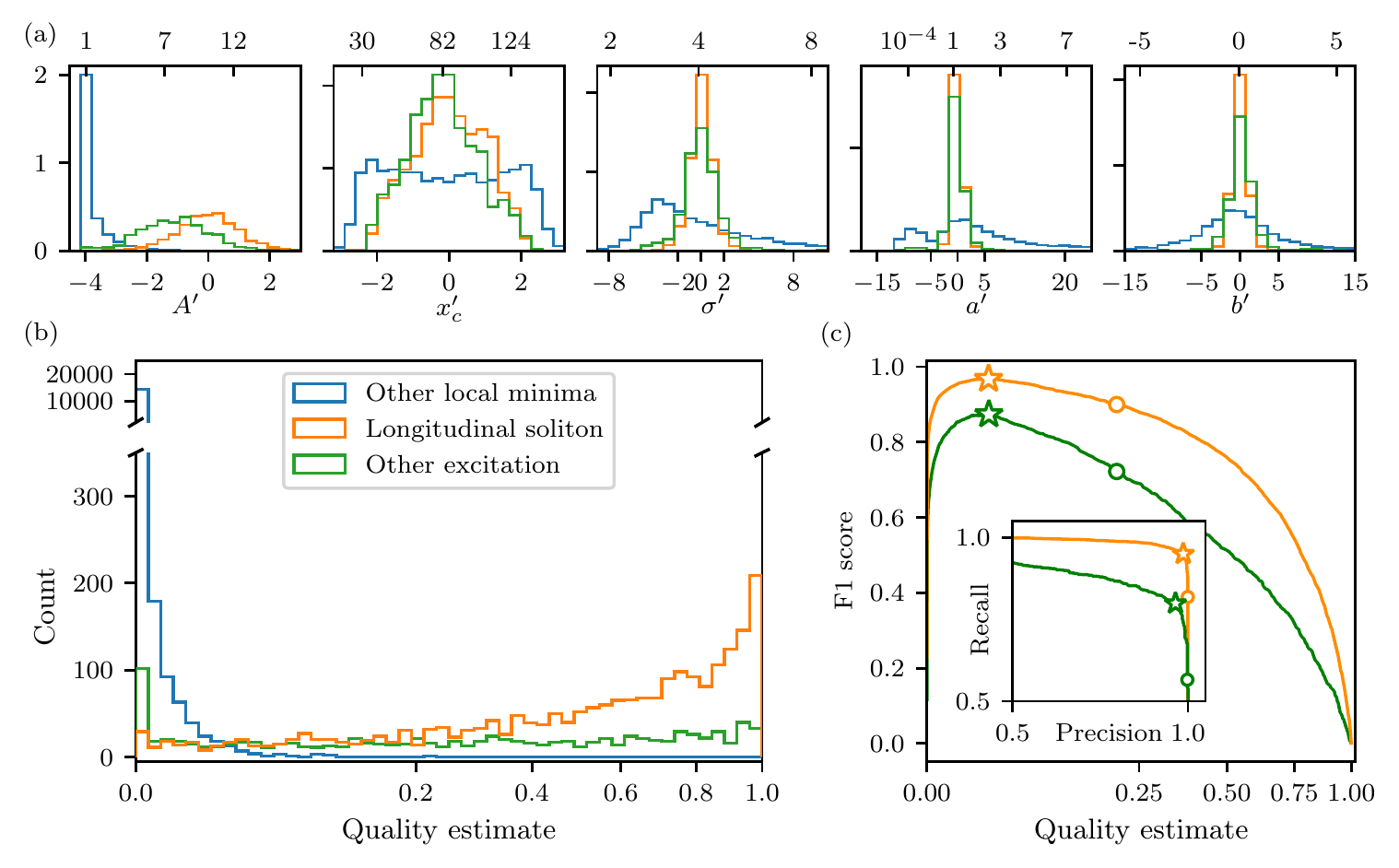}
    \caption{
    Quality estimate performance on no excitation and single excitation classes.
    In all cases we use the following color scheme: longitudinal solitons (orange), all other solitonic excitations (green), and all nonsolitonic local minima (blue).
    (a) Power transformed fit coefficient distributions, with untransformed variables labeled on the top axis. 
    (b) Distribution of quality estimate.
    (c) Performance of quality estimate quantified by F1 score; the stars indicate the optimal F1 value and the circles mark the threshold we use for classification.
    The inset shows precision/recall curves.
}
    \label{fig:quality_estimate_single}
\end{figure*}

\begin{description}
\item[{\it $A$ cut}] The amplitude parameters $A^\pm$, and their ratio $\rho_A$ allow us to identify excitations that do not span the whole cloud.
Data with $\rho_A > 1.57$ are classified as ``top partial excitation'' and those with $1/\rho_A > 1.57$ are classified as ``bottom partial excitation.''
This threshold identifies large fractional jumps in depth between the top and bottom that likely are off-axis solitonic vortices.
Applying $A$ cuts first is important because partial excitations interfere with the subsequent steps. 
\item[{\it $\delta b$ cut}] Figure~\ref{fig:sample_data}(b)(iii) illustrates a case with large shoulder height difference $\delta b$;  Ref.~\cite{donadello_observation_2014} showed that such data result from solitonic vortices.
As a result, we classify data with $\delta b > 0.75$ as ``counterclockwise solitonic vortex'' and $\delta b < -0.53$ as ``clockwise solitonic vortex.''
\item[{\it $\delta i_c$ cut}] Since longitudinal solitons have a vertically aligned density depletion~\footnote{Our images are slightly rotated with respect to the vertical and horizontal axes leading to a small angle in our data.}, we classify data with  $-3.0 < \delta i_c < 1.14$  as ``longitudinal soliton.''
\item[{\it Weaker $\delta b$ cut}] Figure~\ref{fig:parameters} shows that differences $\delta i_c$ and  $\delta b = b^+/\sigma^+ - b^-/\sigma^-$ are anticorrelated, indicating that asymmetries in position and shoulder height are related.
A closer look at Fig.~\ref{fig:sample_data}(b)(iii) indicates that it is such a case, with $\delta i_c < 0$ and $\delta b > 0$.
We therefore add images with $\delta i_c < -3.0$ and $\delta b > 0.61$ to the counterclockwise solitonic vortex class and those with $\delta i_c > 1.14$ and  $\delta b < -0.41$ to the clockwise solitonic vortex class.
\item[{\it Other data}] The remaining images have $\delta i_c \neq 0$ but $\delta b \approx 0$ are labeled as ``canted excitations,'' likely kink solitons in the process of decay.
\end{description}

The flow chart in Fig.~\ref{fig:classification_alg} shows the application of this classifier to a single-soliton dataset.
We found that of the initial 3\,212 images, about 1/3 failed a cut and were rejected as longitudinal soliton candidates.

This classification was also used in the preparation of Ref.~\cite{Fritsch21-DSD} in which we present a refined soliton dataset, which includes improved single longitudinal soliton labels.
The cuts above are fairly aggressive to avoid false positives in the longitudinal soliton classification.
This implies possible misclassification in the other categories in order to ensure a high quality longitudinal soliton subset and a reliability of the quality metric.

\subsection{Quality estimator}\label{ssec:validation}

The quality estimator is initialized on the subset of the single excitation class identified as longitudinal soliton using the PIE classifier.
Figure~\ref{fig:quality_estimate_single}(a) shows the power-transformed distribution of Mexican-hat fit coefficients $\Theta^\prime$, with nontransformed coordinates marked on the top axis for reference.
As would be expected, the data from the initialization dataset (orange) are nearly normally distributed; interestingly, the remaining elements of the single excitation class (partial solitons, canted excitations, and solitonic vortices, as labeled by the PIE filter) collectively follow very similar distributions (green). 
By contrast, the coefficients from every local minimum~\footnote{We require that the local minima be at least 7 pixels wide, i.e., a minimum at pixel $i$ must obey $n^{\rm 1D}_{i\pm j} < n^{\rm 1D}_{i\pm(j+1)}$ for $j=0,1,2$.} in the initialization set {\it except} solitonic excitations (blue curve) obey a qualitatively different distribution.

\begin{figure*}[t]
    \centering
    \includegraphics[width=0.99\linewidth]{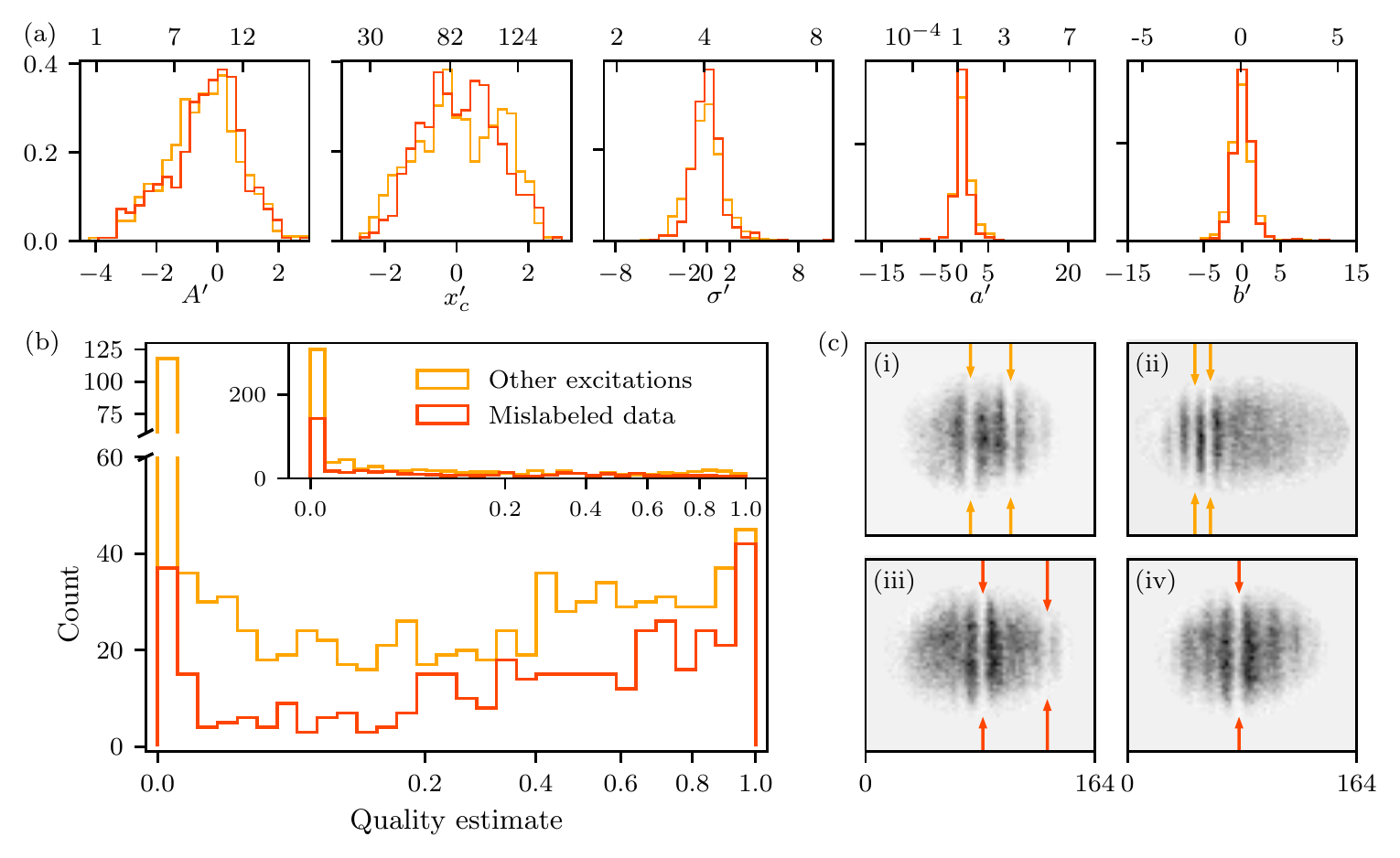}
    \caption{Performance of quality estimate on other excitation (orange) and mislabeled (red) classes.
    (a) Power-transformed fit coefficient distributions, with untransformed variables labeled on the top axis. 
    (b) Distribution of quality estimate of all longitudinal solitons.
    (c) Representative images from the other excitation [(i),(ii)] and mislabeled [(iii),(iv)] classes, with OD+PIE identified longitudinal soliton marked with arrows.
    The quality estimate for these excitations is as follows : $M_{({\rm i})}=[0.74, 0.86]$, $M_{({\rm ii})}=[0.00, 0.01
    ]$, $M_{({\rm iii})}=[0.92, 0.02]$ (all from left to right), and $M_{({\rm iv})}=0.82$. 
}
    \label{fig:objdet_other}
\end{figure*}

Using this initialization, we compare quality estimates $M$ obtained from the single excitation class in Fig.~\ref{fig:quality_estimate_single}(b).
The orange data show $M$ for longitudinal solitons, and as intended the majority of this data is associated with larger values of $M$.
The green data for the remaining solitonic excitations are nearly uniformly distributed, and the nonsoliton minima (blue) are highly concentrated at small $M$.
We note that the small peak in longitudinal soliton distribution near-zero $M$ contains a negligible fraction of the longitudinal soliton dataset (about $1.3~\%$).
However, this peak is more pronounced for the remaining excitations, which is not surprising because the power transform was initialized using longitudinal soliton data.
These distributions demonstrate the ability of the quality estimator to discriminate between solitonic excitations and other features in the data, reinforcing the importance of the PIE filter for fine-grained classification.

We quantify the performance of the quality estimator in terms of the F1 scores plotted in Fig.~\ref{fig:quality_estimate_single}(c), for longitudinal solitons (orange) and all other solitonic excitations (green).
The F1 score for longitudinal solitons is maximized with a threshold of just $M = 0.02$ (stars); however, in practice we minimized false positives and assign a feature to be solitonic when $M > 0.2$ (circles).
This choice gives only small change in the F1 score; however, it gives a marked increase in precision with only a small reduction in recall, as shown in the inset.
The performance of the quality estimate on the other solitonic excitations, while far better than random, is subpar; this reemphasizes the importance of the PIE classifier in our framework.

\subsection{Application to other excitation and mislabeled data class}\label{ssec:test}

Here we discuss the performance of our \textsc{soldet} framework applied to two classes of data from the dark soliton dataset: other excitations (1\,036 images) and mislabeled data (879 images).
These classes consist of images with multiple solitonic excitations, such as shown in Fig.~\ref{fig:sample_data}(c), as well as confusing structures that made human annotation difficult.
As such, they are an ideal test dataset since they defeated previous labeling attempts.

As a reminder, after the CNN classification step, the framework first uses the OD to locate all soliton candidates that are then sorted by the PIE classifier.
Here, we focus only on features identified as longitudinal solitons.
Figure~\ref{fig:objdet_other}(a) plots the frequency of transformed Mexican hat fit outcomes $\Theta^\prime$, giving distributions that for both classes are qualitatively the same as those in Fig.~\ref{fig:quality_estimate_single}(a) for the labeled single solitons.
By contrast, histograms of the quality estimate for longitudinal solitons detected in these two classes [panel (b)] have important differences.
For the other excitations class ($N_{\rm longitudinal}=877$, $N_{\rm images}=669$), the distribution is nearly uniform, with a potential increase for the higher quality estimates ($M>0.4$).
For the mislabeled data ($N_{\rm longitudinal}=415$, $N_{\rm images}=398$), on the other hand, the quality estimate distribution follows a trend consistent with that observed in Fig.~\ref{fig:quality_estimate_single}(b).

To better understand this discrepancy it is important to consider more carefully the differences between the two classes. 
According to the OD module, nearly $78~\%$ of images in the other excitation class contains two or more excitations.
While for excitation spaced apart within the BEC, as in Fig.~\ref{fig:objdet_other}(c)(i), the individual fits to Mexican hat do not affected one another, the contrary holds for excitation captured in close proximity, as shown in Fig.~\ref{fig:objdet_other}(c)(ii).
Qualitative differences between these images are quantified by the quality estimate.
The quality estimates for the two well separated excitation in image (i) is $0.74$ and $0.86$.
In image (ii), in contrast, even though both excitations are reminiscent of a longitudinal soliton, they are assigned a low quality, with $M_{({\rm ii})}=[0.00, 0.01]$ from left to right.
This is likely because the overlap in the adjacent shoulders significantly affects the relative fits.
Given that the majority of data in this class contains multiple excitations, the unusually high frequency of the low quality is to be expected.

The mislabeled class, on the other hand, consists of images determined to be potentially mislabeled during the manual annotation (see Ref.~\cite{Fritsch21-DSD} for details about the data curation process).
These include over 320 images that the annotators found confusing (but in which ODs consistently found exactly one candidate excitation); over 190 images removed during curation from the single excitation class; and about 30 images originally assigned to the no excitation class (but in which the ODs also consistently found exactly one candidate excitation).
Unsurprisingly, in almost $83~\%$ of these images the OD module found only one excitation.
Two representative images from this set are shown in Figs.~\ref{fig:objdet_other}(c)(iii) and \ref{fig:objdet_other}(c)(iv), with $M_{({\rm iii})}=[0.92, 0.02]$ and $M_{({\rm iv})}=0.82$.
The distribution of non-longitudinal soliton quality estimate, shown in the inset in Fig.~\ref{fig:objdet_other}(b), is consistent with that depicted in Fig.~\ref{fig:quality_estimate_single}(b).

The performance on these qualitatively different test sets emphasizes the power of \textsc{soldet}.
By combining the CNN and OD modules, \textsc{soldet} autonomously and reliably locates multiple excitations within the BECs, which goes beyond the traditional state-of-the-art deepest-depletion-based approach.
The PIE classifier enables further systematic validation that the desired type of excitation (here, longitudinal solitons) has been observed, which previously required visual inspection of each acquired image.
Finally, the quality metric provides a quantitative assessment of the excitation quality, further reinforcing the classification reliability.
Put together, these tools provide a robust and reliable analysis framework, capable of processing data significantly more complex than possible given the current traditional state-of-the-art approaches.

\subsection{{\scshape soldet}: Open-source \textsc{python} package for solitonic excitation detection}\label{ssec:soldet}

\begin{figure}[t]
    \centering
    \includegraphics{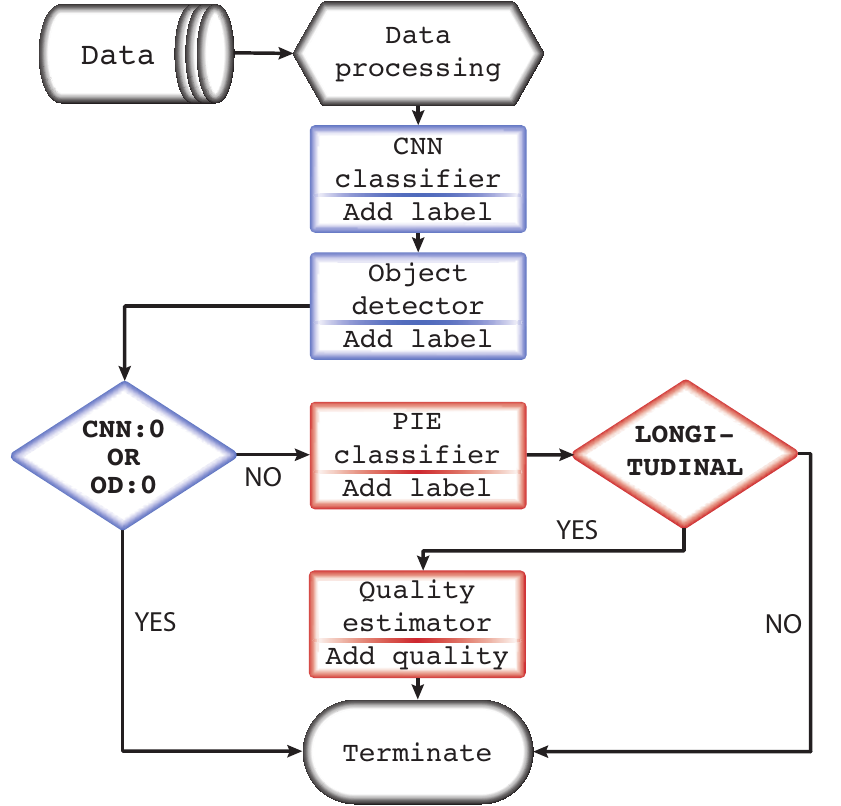}
    \caption{The \textsc{soldet} flow chart.
    The black line follows the \textsc{soldet} dataflow and contains the labels added by each module (rectangles). 
    Blue blocks represent ML modules; red blocks represent physics-based modules.
    }
    \label{fig:soldet}
\end{figure}

In this section, we describe our software package \textsc{soldet} that integrates both the ML modules (CNN classifier and OD) with the fitting physics-based modules (PIE classifier and quality estimator), as we described in previous sections. 
The above discussion showed that the ML modules classify images effectively and can accurately locate one or many candidate solitons.
The physics-based modules can sort these candidates into subclasses and provide a quality estimate for longitudinal soliton candidates.
Therefore, the ML and physics-based modules contribute to the task of soliton detection in different ways, and the \textsc{soldet} infrastructure leverages their complementing strengths. 
We emphasize that soliton detection is one of a larger class of feature identification in quantum gases and that \textsc{soldet} was designed to be broadly applicable.

The \textsc{soldet} distribution includes a CNN classifier, OD, PIE classifier, and quality estimator trained and initialized using the soliton dataset~\cite{solitons-data}.  
In addition, we provide training scripts to enable the ready application to user-defined data with custom preprocessors, ML models, fitting functions, and even the overall process flow.

Figure~\ref{fig:soldet} illustrates a single use of \textsc{soldet} for the specific example of longitudinal soliton detection, where the individual blocks operate as follows: 
\begin{description}
\item[{\it Data processing}] Preprocess raw data into a $164 \times 132$ image format that just encloses the elliptical atom clouds~\cite{guo_machine-learning_2021}.
The preprocessing particulars are not generic and instead are specific to both our task as well as the experimental parameters.
\item[{\it CNN classifier}] Apply a trained CNN classifier to processed data and yield labels no excitation, single excitation, or other excitations.
\item[{\it Object detection}] Apply trained OD to processed data and yield a list of positions of solitonic excitations.
\item[{\it CNN:0 OR OD:0}] If either the CNN classifier or OD finds no soliton, \textsc{soldet} terminates. 
\item[{\it PIE classifier}] The PIE classifier is applied to each solitonic excitation.
\item[{\it Quality estimator}] The quality estimator is applied to each excitation identified as ``longitudinal soliton'' by the PIE classifier.
\end{description}
This algorithm is designed to be usable in a laboratory environment where one needs real-time identification, as well as for automated labeling of large datasets, as in Ref.~\cite{Fritsch21-DSD}. 

\section{Discussion and Outlook}\label{sec:discussion}

Here we described a framework that adds to the growing ML quantum science and technology toolkit, with additional recent developments including noise characterization~\cite{harper_efficient_2020,Ziegler22-TRA}; quantum state detection~\cite{carrasquilla_machine_2017,zhang_quantum_2017,torlai_neural-network_2018,venderley_machine_2018,rem_identifying_2019, miles_correlator_2020, cha_attention-based_2020,metz_deep_2020,venderley_harnessing_2021,maskara_learning_2021, guo_machine-learning_2021}; parameter space exploration and optimization~\cite{wigley_fast_2016,tranter_multiparameter_2018,kalantre_machine_2019,zwolak_autotuning_2020,barker_applying_2020,saito_creation_2020}; and quantum control~\cite{baum_experimental_2021, ai_experimentally_2021}. 
Together, these results show that ML techniques can extract information from ambiguous data, efficiently search large parameter spaces, and optimally control quantum systems.

Our high level framework combines ML methods with physics-based analysis, providing an integrated platform for studying experimental data.
Our implementation of this framework, \textsc{soldet}, currently targets the identification, classification, and tracking of features in image data generated by cold-atom experiments.
We demonstrated its initialization and performance using a publicly available dark soliton dataset~\cite{Fritsch21-DSD}.
This investigation focused only on properties of individual images; however, the dataset also includes a label giving time elapsed since the excitation's were created.
This opens the door for studies correlating system control parameters and the \textsc{soldet} labels.

While our initialization used only the no excitation and single excitation classes, \textsc{soldet}'s feature detection successfully generalizes the learned patterns.
This is confirmed by its performance on the other excitations and mislabeled classes that were not part of training, where the CNN classifier gave ambiguous results and human classifiers often disagreed.
Going beyond simple classification tasks, \textsc{soldet} allowed us to identify unexpected structure in the data, enabling a fine-grained division of the single excitation class into physically relevant subclasses, including solitonic vortices and partial solitons.

Moreover, for the multiple excitations class, the distribution of the quality metric in Fig.~\ref{fig:objdet_other} reveals a possible correlation between the quality metric and the excitations relative proximity. 
These observations illustrate the power of our combined framework as a data analysis tool for discovery. 

An interesting application of \textsc{soldet} would be an off-line optimization of the experimental setup.
Such optimization strategy, successfully implemented to, e.g., improve fabrication of quantum dot devices~\cite{Mei20-OQF}, requires an efficient analysis of large volumes of data to find the appropriate correlations in a high-dimensional parameter space.
The ML toolbox described in our paper allows one to automatically locate multiple solitonic excitations in the same cloud and produces a fine classification that goes beyond longitudinal solitons.
An analysis of the correlations between the various control parameter ranges used in our experiments and the resulting class of data (as determined by \textsc{soldet}) could enable a controlled generation of a desired number, type, and configurations of excitations, with \textsc{soldet} integrated online to provide real-time data analysis and control feedback.
Another interesting extension of this work would be to train an OD on a dataset containing a single subclass found by the PIE classifier, e.g., longitudinal solitons, or solitonic vortices.

From the ML perspective, adding modules based on unsupervised~\cite{celebi2016unsupervised}, active learning~\cite{sun2010survey}, and synthetic data generation with generative models~\cite{gui2020review} may further enhance the performance of the \textsc{soldet} framework.

Going beyond solitonic excitations, the wakefield for sub- and supersonic impurities moving in atomic superfluids have characteristic patterns that could be identified by ML techniques~\cite{PhysRevLett.97.260403, PhysRevA.75.033619, PhysRevA.79.053619, PhysRevLett.100.160402}.
This might be implemented using a template-based method such as used in the Laser Interferometer Gravitational-Wave Observatory (LIGO) where a large set of numerical simulations provide a library of patterns to correlate with the data~\cite{PhysRevD.77.104017}.
This pattern matching is a form of object detection, and in our context a CNN based object detector could also be trained on such a template set.
In this way, our methodology could be employed with a trained OD followed by a LIGO-like algorithm playing the role of our quality estimator and PIE classifier.

In the final analysis, \textsc{soldet} improves the data analysis pipeline for feature identification and classification problems in physically derived image data, but leaves the remainder of the scientific discovery process unchanged.
For example, in our studies the PIE classifier module provided a fresh way to process data and enabled us to identify new patterns in the reduced data.
The step beyond this is ML-driven discovery, where the identification of previously unknown patterns and physical reasoning are both implemented by ML.
An emerging area of ML is the derivation of effective hydrodynamic equations of motion for biological, colloidal, and active fluids based on time-series data~\cite{Supekar2021}.
Owing to the complexity of full 3D simulations of nonzero temperature BECs, this data-driven approach could also be applied to create effective kinetic theory of solitons as well as the hydrodynamics of the underlying fluid.

\section*{Acknowledgements}\label{sec:ackn}
This work was partially supported by NIST and NSF through the Physics Frontier Center at the JQI.
The views and conclusions contained in this paper are those of the authors and should not be interpreted as representing the official policies, either expressed or implied, of the U.S. Government. 
The U.S. Government is authorized to reproduce and distribute reprints for Government purposes notwithstanding any copyright noted herein.

\begin{figure}[t]
    \centering
    \includegraphics[width=0.99\linewidth]{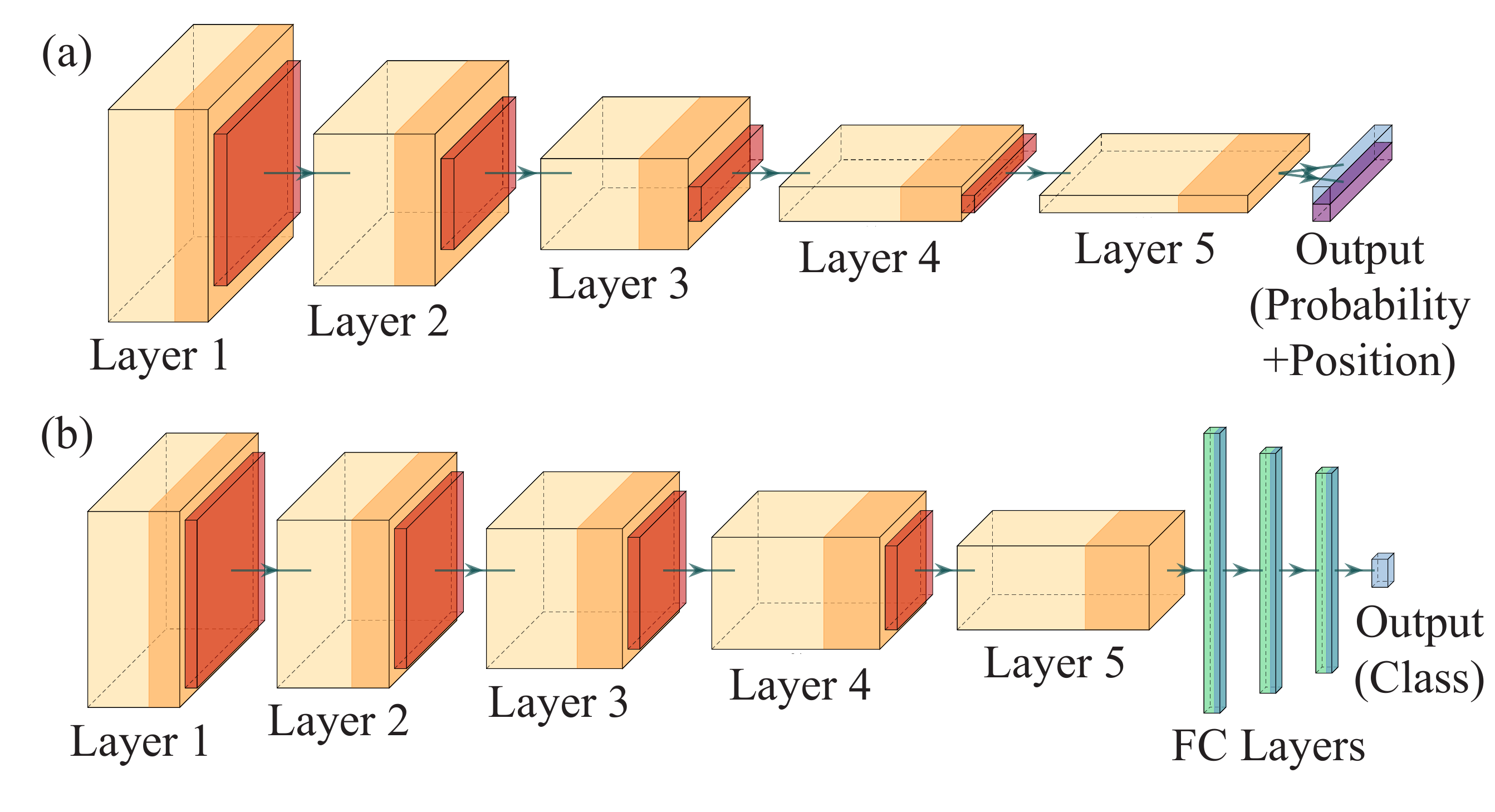}
    \caption{Illustration of (a) OD and (b) CNN classifier neural network architectures. 
    Yellow-orange boxes show convolutional layers while orange-red boxes show max-pooling layers. 
    The horizontal lengths of boxes represent the number of filters and the other two dimensions represent the image sizes. 
    The horizontal blue and purple rectangles in (a) denote output vectors.
    Each cell of the blue vector describes the probability that it contains a soliton and and the purple vector contains the position of a soliton within the cell. 
    And the vertical blue-green rectangles in (b) are three fully connected layers and the output layer.
    The lengths of edges are logarithmically scaled.
}
    \label{fig:pipeline}
\end{figure}

\begin{table}[t]
    \caption{The OD architecture parameters.
    The top four rows are for the  convolutional 2D layers and the three bottom rows are for max-pooling 2D layers.}
    \label{tab:od_architecture}
    \centering
    \begin{tabular}{c|cccccc}
        \hline\hline 
        Layer       & 1 & 2 & 3 & 4 & 5 & Output \\
        \hline 
        Filter & 8 & 16 & 32 & 64 & 128 & 2\\
        Kernel & 5$\times$5 & 5$\times$5 & 5$\times$5 & 1$\times$5 & 1$\times$5 & 1$\times$5\\
        Padding & Same & Same & Same & Same & Same & Same \\
        Activation & ReLu & ReLu & ReLu & ReLu & ReLu & Sigmoid\\
        \hline
        Pool size & 4$\times$2 & 4$\times$2 & 4$\times$1 & 2$\times$1 & N/A & N/A\\ 
        Strides & 4$\times$2 & 4$\times$2 & 4$\times$1 & 2$\times$1 & N/A & N/A\\ 
        Padding & Valid & Valid & Same & Same & N/A & N/A \\
        \hline\hline
    \end{tabular}
    
\end{table}

\appendix
\section{Parameters of Machine Learning Models}\label{app:ml}
Both machine learning modules are built and trained using the \textsc{tensorflow} (v.2.5.0) \textsc{keras python api}~\cite{tensorflow2015-whitepaper}. 
Figures~\ref{fig:pipeline}(a) and \ref{fig:pipeline}(b) show the visualization of the network architecture for the OD and the CNN classifier, respectively. 
The model parameters of OD are presented in Table~\ref{tab:od_architecture}. 
The model parameters for the CNN classifier are presented in the Appendix of Ref.~\cite{guo_machine-learning_2021}.

As can be seen in Fig.~\ref{fig:pipeline}, there are three main differences between the two architectures: (1) the OD outputs 41 local probabilities and positions while the CNN classifier only outputs one of three possible classes;
(2) the CNN classifier contains three fully connected layers, which dramatically increase the number of trainable parameters, while OD does not;
(3) the OD has asymmetric pool size and strides for vertical and horizontal directions, which are customized to the features in our dataset; the pool size and strides are symmetric for the CNN classifier.
As a result, the OD has more than an order of magnitude fewer trainable parameters ($7\times10^4$) than the CNN classifier ($10^6$).


\end{document}